\documentclass[aps,prd,twocolumn,showpacs,groupedaddress,nofootinbib]{revtex4-1}
\usepackage{dcolumn}   
\usepackage{bm}        
\usepackage[english]{babel}
\usepackage{amsfonts,amsmath,amssymb,mathrsfs}
\usepackage{epstopdf}
\usepackage{subfigure}
\usepackage{color}
\usepackage[hidelinks]{hyperref}
\usepackage{hyperref}
\usepackage[pdftex]{graphicx}

\def\a{\alpha}
\def\r{\rho}
\def\s{\sigma}
\def\t{\tau}
\def\m{\mu}
\def\n{\nu}
\def\k{\kappa}
\def\th{\theta}
\def\g{\gamma}\def\G{\Gamma}
\def\L{\Lambda}\def\l{\lambda}
\def\D{\Delta}
\def\la{\langle}
\def\ra{\rangle}
\def\o{\omega}\def\O{\Omega}
\def\d{\delta}
\def\p{\partial}

\def\half{\textstyle{\frac{1}{2}}}
\def\third{\textstyle{\frac13}}

\def\bdoc{\begin{document}}
\def\edoc{\end{document}}

\def\beq{\begin{equation}}
\def\eeq{\end{equation}}
\def\bea{\begin{eqnarray}}
\def\eea{\end{eqnarray}}
\def\ben{\begin{enumerate}}
\def\een{\end{enumerate}}
\def\la{\langle}
\def\ra{\rangle}
\def\a{\alpha}
\def\b{\beta}
\def\g{\gamma}\def\G{\Gamma}
\def\d{\delta}\def\D{\Delta}
\def\e{\epsilon}
\def\z{\zeta}

\def\th{\theta}
\def\k{\kappa}
\def\l{\lambda}
\def\m{\mu}
\def\n{\nu}
\def\o{\omega}
\def\p{\pi}
\def\r{\rho}
\def\s{\sigma}
\def\t{\tau}
\def\L{{\cal L}}
\def\S{\Sigma }
\def\gsim{\; \raisebox{-.8ex}{$\stackrel{\textstyle >}{\sim}$}\;}
\def\lsim{\; \raisebox{-.8ex}{$\stackrel{\textstyle <}{\sim}$}\;}
\def\gtrsim{\gsim}
\def\lessim{\lsim}
\def\loc{{\rm local}}
\def\vm{v_{\rm max}}
\def\bh{\bar{h}}
\def\del{\partial}
\def\nab{\nabla}
\def\half{{\textstyle{\frac{1}{2}}}}
\def\fourth{{\textstyle{\frac{1}{4}}}}

\def\bD{{\bf D}}
\def\bE{{\bf E}}
\def\bF{{\bf F}}
\def\bB{{\bf B}}
\def\bP{{\bf P}}
\def\bV{{\bf v}}
\def\bv{{\bf v}}
\def\bx{{\bf x}}
\def\by{{\bf y}}
\def\bz{{\bf z}}
\def\ba{{\bf a}}
\def\bd{{\bf d}}
\def\bs{{\bf s}}
\def\bn{{\bf n}}
\def\bp{{\bf p}}

\def\O{\Omega}

\def\br{{\bf r}}
\def\bnab{{\bf \nab}}

\def\tE{\tilde{E}}
\def\tL{\tilde{L}}
\def\Horava{Ho\v{r}ava }

\begin{document}

\title{Undoing the twist: the \Horava limit of Einstein-aether}
\author{Ted Jacobson}
\affiliation{Center for Fundamental Physics,  University of Maryland, College Park, MD 20742-4111, USA.\\
Institut d'Astrophysique de Paris, 98bis bv.\ Arago, 75014 Paris, France.}
\date{\today} 
\begin{abstract}
We show that \Horava gravity can be obtained from Einstein-aether theory
in the limit that the twist coupling constant goes to infinity, while holding
fixed the expansion, shear and acceleration couplings.
This limit helps to clarify the relation between the two theories,
and allows \Horava results to be obtained from Einstein-aether ones. 
The limit is illustrated with several examples, including rotating black hole 
equations, PPN parameters, and radiation rates from binary systems.

\end{abstract}  
\maketitle

\section{Introduction}

Einstein-aether theory 
\cite{Gasperini,Jacobson:2000xp,Jacobson:2008aj} is a generally covariant modification
of general relativity (GR), in which the spacetime metric $g_{ab}$ is coupled to 
a unit timelike vector field $u^a$, the ``aether".
\Horava gravity \cite{Horava:2009uw,Blas:2009qj} 
is a related theory, in which the aether is kinematically restricted
at the level of the action to be twist-free or, equivalently, hypersurface orthogonal. 
Originally formulated in terms of the spatial metric, lapse and shift on a fixed foliation, 
\Horava gravity includes higher spatial derivative
terms leading to ``Lifshitz scaling" at short distances, that render the theory 
power counting renormalizable as a quantum field theory.
It is not clear whether a similar Lifshitz version can be formulated for 
Einstein-aether theory, since it lacks the preferred hypersurfaces that play a role
in defining the scaling behavior. 
Here I will restrict attention to the IR limits of these theories, 
although if a Lifshitz extension of Einstein-aether theory exists, 
the main result of the paper would generalize to that context.

At second order in derivatives,
the Einstein-aether (hereafter called ``\ae-theory") 
Lagrangian is a combination of five scalars (including the Ricci scalar $R$) 
that yield independent 
contributions to the equations of motion. 
The second order lagrangian for \Horava 
gravity (hereafter called ``$T$-theory", and which is also known as Khronometric gravity \cite{Blas:2010hb}) 
consists of the same scalars, although
the hypersurface-orthogonal (HO) condition reduces the number of independent ones
from five to four \cite{Blas:2010hb,Jacobson:2010mx}.  
Any HO solution to the \ae-theory field equations is  
also a $T$-theory solution, since the 
action is stationary with respect to \textit{all} variations of the aether
(preserving unit norm), not just HO variations \cite{Jacobson:2010mx}. 
For the same reason, some $T$-theory solutions
are not \ae-theory solutions. An 
example is provided by the (perturbative) rotating black hole solutions 
\cite{Barausse:2012ny, Wang:2012at, Barausse:2012qh, Wang:2012nv}.

The main purpose of this paper is to identify a limit of \ae-theory that 
coincides with $T$-theory. This limit includes not only the HO solutions 
of \ae-theory, but also $T$-theory solutions
that are not \ae-theory solutions, for example the 
rotating black hole solution just mentioned. I have not quite proved that 
\textit{all} $T$-theory solutions arise via this limit, but
I will argue that, if they do not, it is only because of boundary 
effects. Having established this limit, the relation between the theories
can be better understood, and 
information about $T$-theory can be obtained using 
results previously established for Einstein-aether theory.
 
In brief, the argument goes as follows. 
One of the five independent terms in the Lagrangian is chosen 
to be the square of the twist of the aether (to be defined explicitly below), 
which vanishes if and only if the aether is HO. 
The coupling coefficient of this term is sent to infinity, so that the 
twist terms in the aether and metric equations of motion diverge unless they vanish. 
In this limit, a solution can remain regular only if all of these 
terms vanish; but not all of them can vanish unless the twist
itself vanishes. This limit therefore selects the HO Einstein-aether
solutions. More surprisingly, the limit has the effect of suppressing the terms in the 
field equation that come from non-HO variations of the metric, so 
one obtains also $T$-theory solutions that are \textit{not} \ae-theory solutions.
This phenomenon is explained in Sec.~\ref{Tlimit}, and 
illustrated in Sec.~\ref{RBH} using the rotating black hole solution.
In Sec.~\ref{parameters} various parameters 
characterizing the physics of $T$-theory are derived from
those of \ae-theory by use of this limit, and Sec.~\ref{Obs}
discusses how the difference in PPN parameters affects
the observational constraints. 

\section{\ae-theory and $T$-theory}

The two-derivative action for \ae-theory has the form 
\beq \label{S}
S = \frac{1}{16\pi G_0}\int  (-R + L_{\ae})\sqrt{-g}\, d^{4}x
 \eeq
where $R$ is the 4D Ricci scalar, the aether Lagrangian
$L_{\ae}$ is a sum of four independent scalars quadratic in 
$\nabla_a u^b$, and the unit constraint $g_{ab}u^a u^b=1$
is either implicit or imposed with a Lagrange multiplier term.
 (The conventions are those of \cite{Wald} except for the spacetime 
signature, which is taken as $(+{-}{-}{-})$.)
In most of the literature,
the aether Lagrangian has been expressed as
$-(c_1 L_1 + c_2 L_2+c_3L_3+c_4L_4)$, where the  
$c_i$'s are dimensionless coupling constants,
and 
\bea
L_1&=&(\nabla_a u_b)(\nabla^a u^b)\label{L1}\\
L_2&=&(\nabla_a u^a)^2\label{L2}\\
L_3&=&(\nabla_a u_b)(\nabla^b u^a)\label{L3}\\
L_4&=&(u^m\nabla_m u^a)(u^n\nabla_n u_a).\label{L4}
\eea

It is more revealing, however, to use a 
decomposition of the covariant derivative of the aether into
terms transforming by
irreducible representations of the $SO(3)$ Lorentz subgroup
that leaves the aether invariant,
\beq\label{decomp}
\nabla_a u_b = -\frac13 \theta h_{ab} + \s_{ab} +\o_{ab} + u_a a_b.
\eeq
Taking into account the unit norm of $u^a$, the spatial projection of 
$\nabla_a u_b$ is $\nabla_a u_b - u_a a_b$, where 
$a^a = u^m\nabla_m u^a$ is the acceleration (which is orthogonal to $u^a$).
The expansion term $-\frac13\theta h_{ab}$ and shear $\sigma_{ab}$ are the 
trace and trace-free parts of the symmetric part of this spatial projection,
while the twist $\o_{ab}$ is the anti-symmetric part,
\beq
\o_{ab}=\nabla_{[a} u_{b]} - u_{[a} a_{b]}.
\eeq
(A different quantity that contains essentially the same information,
and is also called the twist, is 
$\tilde\o^a\equiv \epsilon^{abcd} u_b \nabla_c u_d = \epsilon^{abcd} u_b\, \omega_{cd} $.)
The twist vanishes if and only if $u_{[a} \nabla_b u_{c]}$ vanishes.
If the aether is HO, $u_a$ can be written as
$u_a=f\nabla_a g$ for some functions $f$ and $g$, hence
$u_{[a} \nabla_b u_{c]}=f\nabla_{[a}g\nabla_{b}f\nabla_{c]}g=0$, 
so the twist vanishes. That the converse holds
is not so obvious, but it is implied by the Frobenius theorem \cite{Wald}.

The four terms in (\ref{decomp}) 
are mutually orthogonal with respect to the spacetime metric $g_{ab}$, 
the first three have zero contraction with $u^a$ on either index, 
the last three have zero contraction with $g^{ab}$,
so the only quadratic scalars that can be formed
from them using $g_{ab}$ and $u^a$ are just their squares formed with the metric. 
These squares can serve as the scalars defining the Lagrangian.
The action for Einstein-aether theory can thus be written as 
\beq\label{action}
S=\frac{-1}{16\pi G_0}\int (R + \third c_\theta \theta^2 + c_\s \s^2 + c_\o \o^2 + c_a a^2)\sqrt{-g}\, d^{4}x. 
\eeq
The relation between the coupling constants defined here and those that have been
used previously for \ae-theory is found by substitution of the decomposition (\ref{decomp})
in (\ref{L1}-\ref{L4}), which yields
\bea
c_\theta &=& c_1+c_3 +3c_2\\
c_\s &=& c_1 + c_3\\
c_\o &=& c_1 - c_3\\
c_a &=& c_1 + c_4.
\eea
These combinations of coupling
constants have been found to appear in numerous physical quantities 
\cite{Jacobson:2008aj}.

The aether in \ae-theory is unconstrained, other than being a unit vector, 
so it has three kinematic degrees of freedom.
In $T$-theory the aether is restricted to be 
HO, so it can be expressed in terms of the gradient of
a scalar field $T$ (the ``khronon" \cite{Blas:2010hb}), via
\beq \label{khronaether}
u^a = \frac{\nabla^a T}{\sqrt{(\nabla_m T)(\nabla^m T)}}.
\eeq
(Note that $u^a$ is unchanged if $T$ is replaced by any 
monotonically increasing function of $T$.)
The action for $T$-theory is the same as the \ae-theory action 
(\ref{action}), but with the aether defined by (\ref{khronaether}) \cite{Blas:2010hb,Jacobson:2010mx}. 
The twist term vanishes identically for HO $u^a$, however,
so the coupling constant $c_\o$ plays no role. A common notation
for the remaining constants $c_a$, $c_\s$, and $c_\theta$ in $T$-theory 
is $\a$, $\b$, and $\b+ 3\l$, respectively \cite{Blas:2011zd}.
 Ho\v{r}ava's original formulation results by choosing $T$ as a spacetime coordinate, thus
fixing some of the coordinate freedom. This coordinate choice 
may be imposed in the action, since the $T$ equation of motion 
is implied by the other equations of motion \cite{Jacobson:2010mx}.

\subsection{Remarks on the form of the action}

As an aside, it is worth making a few remarks about the action (\ref{action}).
This action was first constructed 
in Ref.~\cite{ArmendarizPicon:2010mz} using the irreducible 
parts of the covariant derivative of the Goldstone bosons
associated with the breaking of local Lorentz invariance
down to the rotation group. The authors 
also noted that the coupling constants of the irreducible
terms are the ones that appear more simply in many physical 
properties of the theory. Indeed, with the action expressed
in this way, it is sometimes easy to see which terms, and hence which coupling
constants, can play a role in a given setting. For instance, in a static configuration,
the aether when tangent to a timelike Killing vector has vanishing expansion,
shear, and twist. Since these quantities appear quadratically in the action, those terms
cannot contribute, so only the acceleration coupling 
$c_a$ enters such solutions. As the Newtonian limit
is based on such solutions, the value of Newton's constant $G_N$ (\ref{GN}) 
also depends only on $c_a$. Similarly, in homogeneous, isotropic configurations the aether
has vanishing shear, twist, and acceleration, so such solutions,
as well as the cosmological gravitational constant $G_{\rm cosmo}$ (\ref{Gc}),
involve only the expansion coupling $c_\theta$.
As to waves, the spin (helicity) content of the various quantities is: expansion (0), 
shear (2,1,0), twist (1), and acceleration (1,0). Hence the spin-2 wave speed 
involves only $c_\s$, the spin-1 wave speed involves all but $c_\theta$, and 
the spin-0 wave speed involves all but $c_\o$.

\section{$T$-theory limit of \ae-theory}
\label{Tlimit}

Now consider the $c_\o\rightarrow\infty$ limit of \ae-theory, with all other 
coupling constants held fixed.\footnote{A similar method was used in Ref.~\cite{Blas:2009qj}
to relate the projectable case of \Horava gravity (lapse function spatially constant) 
to the non-projectable case, taking the limit $c_a\rightarrow\infty$.}
(In terms of the usual parameters,
this limit corresponds to $c_1-c_3\rightarrow\infty$, with $c_1+c_3$, $c_1+c_4$, and 
$c_2$ held fixed. Thus $c_1$, $c_3$, and $c_4$ all diverge in this limit.)
Any term in the equations of motion
involving $c_\o$ will diverge in this limit unless that term vanishes. One
of these terms, which arises from varying the volume element, 
is $\sim c_\o \o^2 g_{ab}$, and the only way this can vanish is if 
$\o_{ab}=0$. Thus we may conclude that only HO solutions
are regular in this limit. However, 
if the limit is to coincide with $T$-theory, then not only \ae-theory
solutions but \textit{all} $T$-theory solutions should arise in the limit.
This would be the case only if the limit somehow ``turned off" the
part of the aether field equations that arises from non-HO variations of the action.
This seems indeed to be the case, except for possible boundary effects. 

To understand how this works, 
it is instructive to consider an elementary analogy, in which the 
action $S(x,y)$ depends on just two variables $x$ and $y$, with
$y$ playing the role of the twist.
Expanding in $y$, we have
\beq
S(x,y) = S_0(x) + S_1(x)y+ c_y\, S_2(x)y^2+\dots,
\eeq
where $c_y$ is a coupling parameter, analogous to $c_\o$, 
that we will take to infinity.
Dropping for the moment any higher order terms in $y$,
the $y$ and $x$ equations of motion are given, respectively, by 
\bea
S_1(x) + 2c_y\, S_2(x) y   &=& 0\label{yeq}\\
S_0'(x) +S_1'(x) y + c_y\, S_2'(x) y^2 &=& 0.\label{xeq}
\eea
Unless $S_2(x)$ vanishes, the only regular solutions
in the limit $c_y\rightarrow\infty$ 
will have $y=0$. If we set $y$ to zero before taking the limit,
then the $y$ equation \eqref{yeq} survives as $S_1(x)=0$. 
However, we could also solve the $y$ equation for every finite 
$c_y$, obtaining $y = -S_1(x)/(2c_y S_2(x))$. 
In the limit, we then have $y=0$, and the $y$ equation has been satisfied 
without imposing any further conditions on $x$. 
If $S(x,y)$ contains higher order terms in $y$ that are
independent of $c_y$, then still the only regular solutions
in the limit will be those with $y=0$, and still the $y$ equation
will imply nothing about $x$.

This simple example can be generalized to any finite 
number of coordinates, and even to a field theory setting. 
However, when the variables are fields and the action involves 
derivatives, integration by parts can lead to 
boundary terms. Boundary terms could be avoided by 
working on a compact space, or by fixing boundary conditions. 
I therefore expect the argument should hold locally for \ae-theory,
although in a non-compact space some $T$-theory solutions might 
perhaps be missing from the limit. 

\subsection{Rotating black hole}
\label{RBH}
An example of the appearance, in the limit, of 
$T$-theory solutions that are not \ae-theory solutions, occurs in
the case of the slowly rotating black holes. There are 
no HO \ae-theory solutions \cite{Barausse:2012ny, Barausse:2012qh}, 
but there are $T$-theory solutions
of this type \cite{Barausse:2012ny, Wang:2012at, Barausse:2012qh, Wang:2012nv}. 
We can see them arise in the $c_\o\rightarrow\infty$ 
limit, making use of the field equations in Ref.~\cite{Barausse:2013nwa},
which include the first order contributions of the rotation.

Consider in particular equations (60) and (61) of that 
paper, which hold in \ae-theory 
under the assumption that the spacetime is asymptotically flat:
\bea
d_1\psi' + d_2\psi'' +d_3\l' + d_4\l''&=& 0\label{psieqn-d}\\
b_1\psi' + b_2\psi'' + b_3\l' + b_4\l''  &=& -r^{-4}\l,\label{psieqn-b}
\eea
where $\psi$, $\l$ and all the coefficients are functions of $r$.
The coefficient functions 
depend on the background spherical black
hole solution and the coupling constants.
The function $\psi(r)$ determines the angular velocity of the aether, 
while the function $\l(r)$ governs the twist, in the sense that 
the twist vanishes if and only if $\l(r)=0$. 
It has been shown that
asymptotic flatness requires $\l\rightarrow0$ as
$r\rightarrow\infty$ \cite{bscommunication}.

If the aether is HO, so that $\l(r)=0$, then by 
combining the two equations
one finds that $\psi'(r)=0$, which implies that 
the solution must be the non-rotating black hole 
(generally  
in rotating coordinates) \cite{Barausse:2012qh}. 
That is, there is no 
HO rotating black hole solution in \ae-theory.
In $T$-theory, on the other hand, Eq.~(\ref{psieqn-d})
is not present, and the general solution to Eq.~(\ref{psieqn-b})  
for $\psi(r)$ describes
a rotating black hole.

Consider now what happens if $\l(r)$ is 
kept non-zero in the \ae-theory equations as the limit is taken. 
The functions $b_{1,2,3,4}$ are independent of $c_\o$, 
the functions $d_{1,2,4}$ diverge as $c_\o$ in the limit, 
and the function $d_3$ diverges as $c_\o^2$.
Therefore (\ref{psieqn-d}) can be viewed as an equation for 
$\l(r)$ which implies that $\l'(r)=0$ in the limit. Since all asymptotically
flat \ae-theory solutions have $\l(\infty)=0$, this implies  
$\l(r)=0$. The remaining equation thus reduces in the
limit to the $T$-theory equation,
and so the $T$-theory solutions
are recovered in the limit. 

\subsection{$T$-theory parameters from \ae-theory ones}
\label{parameters}
Quantities characterizing $T$-theory can be obtained as
the infinite $c_\o$ limit of those characterizing \ae-theory.
In this section a number of such quantities will be examined,
both as an illustration of the correspondence, and 
to elucidate the relation between the two theories.

To begin with, the wave speeds  
$s_{2,1,0}$ for the 
spin-2, spin-1, and spin-0 waves in \ae-theory are given by \cite{Jacobson:2004ts} 
\bea
s_2^2 &=&  \frac{1}{1-c_\s} \\ 
s_1^2 &=&  \frac{c_\s + c_\o-c_\s c_\o)}{2c_a(1-c_\s)} \label{s1}\\ 
s_0^2 &=&  \frac{(c_\theta+2c_\s)(1-c_a/2)}{3c_a(1-c_\s)(1+c_\theta/2)}.  \label{s0}
\eea
The spin-2 and spin-0 wave speeds 
are independent of $c_\o$ so they 
coincide in the two theories, while the spin-1 wave
speed diverges as $\sqrt{c_\o}$. 
The energy density at fixed amplitude and frequency 
remains finite \cite{Eling:2005zq,Foster:2006az}, so the energy flux from a periodic source
behaves as $\sqrt{c_\o}$ times the squared wave amplitude.
One can see by inspection
of the radiation analysis of Ref.~\cite{Foster:2006az} that the wave amplitude generated
by such a source vanishes as $c_\o^{-3/2}$, so we conclude that the power radiated 
in spin-1 waves vanishes as $c_\o^{-5/2}$. The spin-1 mode thus
decouples in this limit. 

Newton's constant $G_N$ and the cosmological gravitational 
constant $G_{\rm cosmo}$ appearing in the Friedmann equation
are related to the ``bare" constant $G$ in the 
action (\ref{action}) by  \cite{Carroll:2004ai}
\bea
G_N&=&G/(1-c_a/2)\label{GN}\\ 
G_{\rm cosmo}&=&G/(1 + c_\theta/2).\label{Gc}
\eea
Both of these relations are independent of 
$c_\o$, hence they hold also in $T$-theory, as shown independently in Ref.~\cite{Blas:2009qj}.

The PPN parameters are also related by this limit. All PPN parameters of ae-theory 
are identical to those of GR, except for the preferred frame parameters $\a_1$
and $\a_2$, which are given by \cite{Graesser:2005bg,Foster:2005dk}
\bea
\a_1 &=& 4\frac{c_\o(c_a - 2c_\s)  + c_a c_\s}{c_\o(c_\s-1) - c_\s}\label{a1}\\
\a_2 &=& \frac{\a_1}{2}+\frac{3(c_a - 2c_\s) (c_\theta + c_a)}{(2-c_a)(c_\theta +2 c_\s)}. \label{a2} 
\eea
In the infinite $c_\o$ limit $\a_1$ becomes
\beq\label{a1T}
\a_1^T = 4\frac{c_a - 2c_\s}{c_\s-1},
\eeq
and the formula (\ref{a2}) for $\a_2$ is otherwise unchanged,
in agreement with previous computations for $T$-theory \cite{Blas:2010hb,Blas:2011zd}. 

Finally, radiation amplitudes and rates can be carried over in the limit from 
\ae-theory to $T$-theory, both in the weak and strong self-gravity 
regimes. For weak self-gravity, the results of \cite{Blas:2011zd} are obtained
from the limit of those of \cite{Foster:2006az} (modulo typos), while for strong
self-gravity, the limit of \cite{Foster:2007gr} should agree with 
new results for $T$-theory, 
which have recently been derived \textit{ab initio} in \cite{Yagi:2013ava} and 
reported in \cite{Yagi:2013qpa}. (In fact, computational errors in
\cite{Foster:2007gr} were discovered by the failure of the limit to agree.)

\subsection{Observational constraints}
\label{Obs}

Equations (\ref{a1}-\ref{a1T}) reveal an important discrepancy in how 
$\a_1$ and $\a_2$ can be set to zero (or to small values) in the two theories \cite{Blas:2011zd}. 
In \ae-theory two conditions on the parameters are required \cite{Foster:2005dk}:
\bea
c_a &=& 2\, \frac{c_\o c_\s}{c_\s+c_\o}\longrightarrow 2c_\s\label{PPN1}\\
c_\theta &=& -c_a= -2\frac{c_\o c_\s}{c_\o+c_\s}\longrightarrow  -{2c_\s},\label{PPN2}
\eea
where the arrows denote the limit $c_\o\rightarrow\infty$.
(An alternative would be to set $c_a=0=c_\s$, which would make the spin-0 and 
spin-1 wave speeds diverge.)
In $T$-theory the first condition (\ref{PPN1}) alone, $c_a = 2c_\s$, suffices to set both 
$\a^T_1$ and $\a^T_2$ to zero. (Since there is one less coupling constant to begin with in
$T$-theory, this again leaves a two-dimensional coupling constant space unconstrained,
as in \ae-theory.)
Hence the second condition, (\ref{PPN2}), need not be applied,
so the limit of the \ae-theory conditions for the vanishing 
of the PPN parameters is
stronger than the $T$-theory condition. 

On the other hand, this second condition (\ref{PPN2}) for \ae-theory
implies also that $G_N=G_{\rm cosmo}$ (\ref{GN},\ref{Gc}), ensuring
agreement with the primordial nucleosynthesis constraint \cite{Carroll:2004ai}
and constraints from the spectrum of CMB and matter anisotropies \cite{Audren:2013dwa}.
If the PPN constraints are met in $T$-theory just by imposing 
the first condition, $c_a = 2c_\s$, then these cosmological constraints must 
be separately imposed as the requirement that the second condition hold
approximately, up to a deviation smaller than something of order $\sim 0.1$ 
for nucleosynthesis \cite{Blas:2011zd} and $\sim 0.01$ for anisotropies \cite{Audren:2013dwa}.

If \textit{exact}
vanishing of $\a_{1,2}$ is imposed in $T$-theory, then 
 $G_{\rm cosmo}/G_N = (1-c_\s)/(1+c_\theta/2)$, from which
 it follows that the spin-0 mode speed (\ref{s0}) vanishes as
 $G_{\rm cosmo}$ approaches $G_N$. This is ruled out by 
 the vacuum \v{C}erenkov constraint \cite{Elliott:2005va} (arising from the observation
of ultra-high energy cosmic rays) which requires the 
mode speed to be at least that of light (minus a tiny amount). 
If the spin-0 mode speed is adjusted to be unity, i.e.\ the 
minimum value allowed by the \v{C}erenkov constraint, 
then it follows that $G_{\rm cosmo}/G_N = 1-3 c_a/2$, so
the nucleosynthesis constraint $|G_{\rm cosmo}/G_N-1|\lesssim 0.13$ 
implies $c_a\lesssim 0.08$, i.e.\ $c_\s\equiv\beta\lesssim 0.04$. At the other extreme,
as $c_a=2c_\s\rightarrow 0$, the spin-0 mode speed diverges, 
$G_{\rm cosmo}/G_N = 1/(1+c_\theta/2)$, and the nucleosynthesis constraint 
implies $c_\theta\equiv 3\l \lesssim 0.30$ ($c_\theta=3\l$ when $c_\s=0$). These
constraints are consistent with those reported in \cite{Yagi:2013qpa}.


It was found in \cite{Blas:2011zd} that 
a qualitative difference in the scalar radiation for the two theories
arises, again due to the different way the PPN parameters
are set to zero. When restricting to $\a_{1,2}=0$,  
radiation sourced by the second time derivative of the
second monopole moment ($\sim \int d^3r\ddot\rho r^2$) vanishes 
in \ae-theory but not in $T$-theory. 
The coupling to this monopole term is proportional to 
\beq\label{monopole}
\frac{\a_1-2\a_2}{c_a-2c_\s}= \frac{6(c_\theta + c_a)}{(c_a-2)(c_\theta + 2 c_\s)}.
\eeq
As explained above, in
\ae-theory the vanishing of the PPN parameters does not require $c_a=2c_\s$ 
(but does entail $c_a + c_\theta=0$), so the monopole coupling vanishes.
In $T$-theory, the vanishing of the PPN parameters requires
$c_a=2c_\s$ only, so the left hand side of (\ref{monopole}) is $0/0$, 
and the right hand side shows that the coupling survives. 


\acknowledgements

I am grateful to E.~Barausse, D.~Blas, T.~Sotiriou and A.~Speranza 
for helpful discussions, comments, and suggestions
on a draft of this paper, and to the Institut d'Astrophysique de Paris
where its preparation was completed.
This research was stimulated
and informed by the Kavli IPMU workshop, ``Gravity
and Lorentz violations", and 
was supported in part by the NSF under grant No.~PHY-0903572.

\end{document}